\documentclass{emulateapj}
\usepackage{apjfonts}

\newcommand{\Msun}      {\mbox{$\rm\,M_{\mathord\odot}$}}

\begin{document}

\lefthead{Obscured HMXB IGR J16207-5129}
\righthead{Tomsick et al.}

\submitted{Accepted by the Astrophysical Journal}

\def\lsim{\mathrel{\lower .85ex\hbox{\rlap{$\sim$}\raise
.95ex\hbox{$<$} }}}
\def\gsim{\mathrel{\lower .80ex\hbox{\rlap{$\sim$}\raise
.90ex\hbox{$>$} }}}

\title{An {\em XMM-Newton} Spectral and Timing Study of IGR J16207--5129:  
An Obscured and Non-Pulsating HMXB}

\author{John A. Tomsick\altaffilmark{1},
Sylvain Chaty\altaffilmark{2},
Jerome Rodriguez\altaffilmark{2},
Roland Walter\altaffilmark{3},
Philip Kaaret\altaffilmark{4},
Gagik Tovmassian\altaffilmark{5}}

\altaffiltext{1}{Space Sciences Laboratory, 7 Gauss Way, 
University of California, Berkeley, CA 94720-7450, USA
(e-mail: jtomsick@ssl.berkeley.edu)}

\altaffiltext{2}{AIM - Astrophysique Interactions Multi-\'echelles
(UMR 7158 CEA/CNRS/Universit\'e Paris 7 Denis Diderot),
CEA Saclay, DSM/IRFU/Service d'Astrophysique, B\^at. 709,
L'Orme des Merisiers, FR-91 191 Gif-sur-Yvette Cedex, France}

\altaffiltext{3}{INTEGRAL Science Data Centre, Observatoire
de Gen\`eve, Universit\'e de Gen\`eve, Chemin d'Ecogia, 16, 
1290 Versoix, Switzerland}

\altaffiltext{4}{Department of Physics and Astronomy, University of
Iowa, Iowa City, IA 52242, USA}

\altaffiltext{5}{Observatorio Astronomico Nacional SPM, Instituto de
Astronomia, Universidad Nacional Autonoma de Mexico, Ensenada, BC, Mexico}

\begin{abstract}

We report on a $\sim$12 hr {\em XMM-Newton} observation of the supergiant
High-Mass X-ray Binary IGR~J16207--5129.  This is only the second soft
X-ray (0.4--15~keV, in this case) study of the source since it was discovered 
by the {\em INTEGRAL} satellite.  The average energy spectrum is very similar 
to those of neutron star HMXBs, being dominated by a highly absorbed 
power-law component with a photon index of $\Gamma = 1.15^{+0.07}_{-0.05}$.
The spectrum also exhibits a soft excess below $\sim$2~keV and an iron 
K$\alpha$ emission line at $6.39\pm 0.03$~keV.  For the primary
power-law component, the column density is $(1.19^{+0.06}_{-0.05})\times 10^{23}$
cm$^{-2}$, indicating local absorption, likely from the stellar wind, and 
placing IGR~J16207--5129 in the category of obscured IGR HMXBs.  The 
source exhibits a very high level of variability with an rms noise level 
of 64\%$\pm$21\% in the $10^{-4}$ to 0.05~Hz frequency range.  Although
the energy spectrum suggests that the system may harbor a neutron star, 
no pulsations are detected with a 90\% confidence upper limit of
$\sim$2\% in a frequency range from $\sim$$10^{-4}$ to 88~Hz.  We discuss
similarities between IGR~J16207--5129 and other apparently non-pulsating 
HMXBs, including other IGR HMXBs as well as 4U~2206+54 and 4U~1700--377.

\end{abstract}


\section{Introduction}

The hard X-ray imaging by the {\em INTEGRAL} satellite \citep{winkler03} has been 
uncovering a large number of hard X-ray sources.  Since {\em INTEGRAL}'s launch 
in 2002 October, 550 sources have been detected by the IBIS instrument in the 
$\sim$20--50 keV band (based on version 29 of the ``General Reference Catalog'').  
Included in these sources are 236 ``IGR'' sources that were unknown or at least 
not well-studied prior to {\em INTEGRAL}.  An important result from the
{\em INTEGRAL} mission has been the discovery of a relatively large number of
High-Mass X-ray Binaries (HMXBs).  There are 37 IGR sources that have been 
classified as HMXBs (and it should be noted that about 1/3 of the IGR sources
are still unclassified).  The IGR HMXBs are interesting both for the large 
number of new systems as well as the specific properties of these systems.  
These include a new class of ``Supergiant Fast X-ray Transients'' 
\citep{negueruela06} that are HMXBs that can exhibit hard X-ray flares that only 
last for a few hours while the X-ray flux changes by orders of magnitude 
\citep{intzand05,sguera06}.  Many of the IGR HMXBs are also extreme in having 
a high level of obscuration ($N_{\rm H}$$\sim$$10^{23}$--$10^{24}$ cm$^{-2}$) due to 
material local to the source \citep{walter06,chaty08}.  For both the SFXTs and 
the obscured HMXBs, it is thought that a strong stellar wind is at least partially 
responsible for their extreme X-ray properties \citep{fc04,walter06,wz07}.

The 37 IGR HMXBs presumably contain either neutron stars or black holes, but 
the nature of the compact object is only clear for the 12 systems for which 
X-ray pulsations from their neutron stars have been detected.  Eleven of these 
systems have pulse periods ranging from 5 to 1300~s, and the 12th system has an 
unusually long period of $\sim$5900~s \citep{patel04}.  Another X-ray property 
that can be taken as evidence for the presence of a neutron star is a very hard 
X-ray spectrum.  Typically, neutron star HMXBs have X-ray spectra that can be 
modeled with a power-law with a photon index of $\Gamma$$\sim$1 that is 
exponentially cutoff near 10--20~keV \citep{nagase89,lutovinov05a}.  Although
there are only a few known black hole HMXBs (e.g., Cygnus~X-1, LMC~X-1, 
LMC~X-3, M33~X-7), their X-ray spectral properties are similar to the general 
class of black hole X-ray binaries with power-law spectra with 
$\Gamma$$\sim$1.4--2.1 in their hardest spectral state \citep{mr06}.  If black 
hole spectra show a cutoff, it is usually close to 100~keV \citep{grove98}, 
significantly higher than seen for neutron star HMXBs.  However, it should 
be noted that evidence from the shape of the spectral continuum alone is 
usually taken only as an indication of the nature of the compact object.  
For example, HMXBs 4U~1700--37 and 4U~2206+54 both have continuum X-ray 
spectra similar to neutron star HMXBs, but they are considered to be only 
probable neutron star systems because pulsations have not been detected.

In this study, we focus on X-ray observations of IGR~J16207--5129, which was 
discovered in the Norma region of the Galaxy relatively early-on in the 
{\em INTEGRAL} mission \citep{walter04a,tomsick04_munich}.  Although it is
a relatively faint hard X-ray source at $3.3\pm 0.1$ millicrab in the
20--40~keV band \citep{bird07}, it has been consistently detected by 
{\em INTEGRAL} as well as in X-ray follow-up observations by {\em Chandra} 
and {\em XMM-Newton} (this work), indicating that it is a persistent source.
The {\em Chandra} observation provided a sub-arcsecond position that allowed
for the identification of an optical counterpart with $R = 15.38\pm 0.03$ 
and an IR counterpart with $K_{\rm s} = 9.13\pm 0.02$ \citep{tomsick06}.  
Based on the optical/IR Spectral Energy Distribution, \cite{tomsick06} found
that the system has a massive O- or B-type optical companion.  Optical and
IR observations confirmed this and indicate a supergiant nature for the
companion \citep{masetti06v,ns07,rahoui08}, and \cite{rahoui08} estimated
a source distance of $\sim$4.1~kpc.  With further IR spectroscopy, the 
spectral type of the companion was narrowed down to B1~Ia and a source
distance of $6.1^{+8.9}_{-3.5}$~kpc was estimated \citep{nespoli08}.

Our current knowledge about the soft X-ray properties of IGR~J16207--5129
comes from the {\em Chandra} observation that was made in 2005.  This showed
that the source has a hard X-ray spectrum with a power-law photon index of
$\Gamma = 0.5^{+0.6}_{-0.5}$, and the source also exhibited significant 
variability over the 5~ks {\em Chandra} observation \citep{tomsick06}.  
Based on these properties, we selected this target for follow-up
{\em XMM-Newton} observations to obtain an improved X-ray spectrum and 
to search for pulsations that would provide information about the nature of
the compact object.  Here, we present the results of the {\em XMM-Newton}
observation.

\section{{\em XMM-Newton} Observations and Light Curve}

We observed IGR~J16207--5129 with {\em XMM-Newton} during satellite revolution 1329.
The observation (ObsID 0402920201) started on 2007 March 13, 8.27~hr UT and lasted 
for 44~ks.  The EPIC/pn instrument \citep{struder01} accumulated $\sim$0.4--15~keV
photons in ``small window'' mode, giving a 4.4-by-4.4 arcminute$^{2}$ field-of-view 
(FOV) and a time resolution of 5.6718~ms.  The mode used for the 2 EPIC/MOS units 
\citep{turner01} is also called ``small window'' mode, and for MOS, its features 
are a 1.8-by-1.8 arcminute$^{2}$ FOV and a time resolution of 0.3~s.
In addition to the {\em XMM-Newton} data, we downloaded the ``current calibration 
files'' indicated as necessary for this observation by the on-line software tool
{\ttfamily cifbuild}.  For further analysis of the data, we used the {\em XMM-Newton} 
Science Analysis Software (SAS-8.0.0) as well as the XSPEC package for spectral
analysis and IDL for timing analysis. 

We began by using SAS to produce pn and MOS images and found a strong X-ray source
consistent with the {\em Chandra} position of IGR~J16207--5129 \citep{tomsick06}.
This is the only source seen in the pn image, but for MOS, five of the outer CCD
detectors are active, and a few other faint sources are detected.  To produce a pn 
light curve, we used SAS to read the event list produced by the standard data 
pipeline.  We extracted source counts from a circular region with a radius of 
$\sim$$35^{\prime\prime}$, which includes nearly all of the counts from the source.  
For subtracting the background contribution, we used the counts from a rectangular 
region with an area 3.5 times larger than the source region located so that no 
point in the background region comes within $2^{\prime}$ of the source.  In producing 
the light curves, we applied the standard filtering (``FLAG=0'' and ``PATTERN$\leq$4'')
as well as restricting the energy range to 0.4--15~keV.

\begin{figure}
\plotone{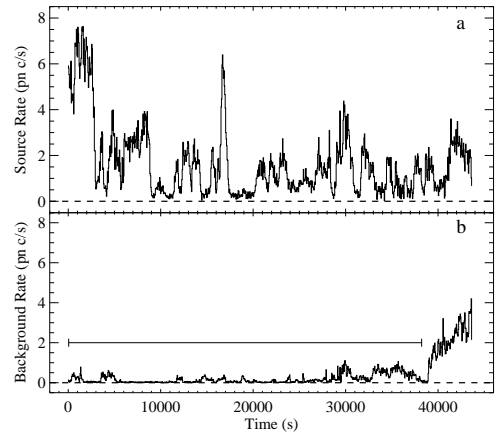}
\caption{(a) {\em XMM-Newton} pn 0.4--15~keV light curve for IGR~J16207--5129.  
The time resolution is 50~s, and we have subtracted the background contribution. 
(b) The light curve from the background region after scaling to the size of the 
source extraction region.  The solid line marks the low-background time segment 
used for spectral analysis.\label{fig:lc}}
\end{figure}

The pn light curves with 50~s time resolution are shown in Figure~\ref{fig:lc}.  
While the average count rate after deadtime correction and background subtraction 
is 1.64~c/s, there is a very high level of variability with count rates in the
50~s time bins ranging from $0.03\pm 0.15$ c/s to $7.6\pm 0.5$ c/s.  The background
light curve after deadtime correction and scaling to the size of the source region
is shown in Figure~\ref{fig:lc}b.  The background rate also shows a high level of
variability during the observation.  The average background rate for the entire 
44~ks observation is 0.40~c/s, but it is much higher at the end of the observation.
For the first 39~ks, the average background rate is 0.18~c/s, while it is 1.9~c/s
for the last 5~ks.  We verified that the higher count rate is due to proton flares 
by producing a pn light curve for the full pn FOV in the $>$10~keV energy band.  
The average full-FOV count rate in this energy band is 0.33~c/s for the first 39~ks 
while it is 4.9~c/s for the last 5~ks.  Thus, for the spectral analysis described 
below, we only included the portion of the exposure indicated in Figure~\ref{fig:lc}.  

We also produced MOS light curves using a circular source region with a radius of
$\sim$$35^{\prime\prime}$, and they show the same variability and flaring as the pn
light curves.  We used the MOS1 and MOS2 pipeline event lists, and applied the 
standard filtering (``FLAG=0'' and ``PATTERN$\leq$12'').  In small window mode, 
the active area of the central CCD is too small to use the data from this CCD for
background subtraction; thus, we used a source-free rectangular region from one of
the outer CCDs as our background region.  We also made MOS light curves using 
data from the full FOVs in the $>$10~keV energy band.  They show that a large 
increase in the background level occurred simultaneously with the increase seen
in the pn, and for the analysis described below, we used the MOS data from the
first 39~ks of the observation (i.e., the low-background time indicated in 
Figure~\ref{fig:lc}b).  After background subtraction, the 0.4--12~keV MOS1 and 
MOS2 count rates in the source region during the low-background time are 0.50 
and 0.51 c/s, respectively.  

The third X-ray instrument on {\em XMM-Newton} is the Reflection Grating 
Spectrometer (RGS).  We inspected the RGS dispersion images that are used to
extract spectra, but there is no evidence that IGR~J16207--5129 is detected.
To determine if an RGS upper limit is constraining, we used the spectra 
obtained using the EPIC instruments (see \S$3.2$ below) along with an RGS
response matrix.  We find that, for the entire 44~ks observation, we would 
expect the RGS to collect $\sim$22 counts and $\sim$13 counts in the first 
and second grating orders, respectively, which, given the instrumental 
background, is consistent with the non-detection.

\clearpage

\section{Analysis and Results}

\subsection{Timing}

We used the {\em XMM-Newton} instrument with the highest effective area, the pn, for 
timing analysis and started with the standard pipeline event list.  We used the SAS 
tool {\ttfamily barycen} to correct the timestamps to the Earth's barycenter for 
each event.  As IGR~J16207--5129 is a HMXB that may contain a pulsar, a primary 
goal of this analysis is to search for periodic signals.  We used the IDL software 
package to read in the event list and make a 0.4--15~keV light curve at the highest 
possible time resolution, $\Delta$$t = 5.6718$~ms.  We note that it is important to 
use this exact value for $\Delta$$t$ to avoid producing artifacts in the power 
spectrum.  Once the high time resolution light curve was produced, we used IDL's 
Fast Fourier Transform (FFT) algorithm to produce a Leahy-normalized power spectrum 
\citep{leahy83}.  The power spectrum extends from $2.3\times 10^{-5}$ Hz (based on 
the 44~ks duration of the observation) to 88~Hz (the Nyquist frequency).  

\begin{figure}
\plotone{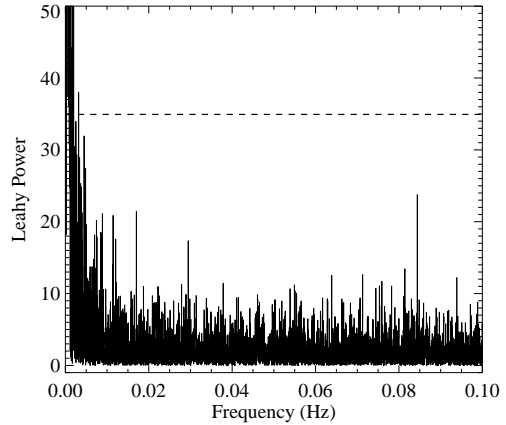}
\caption{The lowest frequency portion of the Leahy-normalized power spectrum 
for IGR~J16207--5129 with a frequency binsize of $2.3\times 10^{-5}$~Hz.  The
horizontal dashed line marks the 90\% confidence detection limit (after
accounting for trials).  The full power spectrum extends to the Nyquist
frequency of 88 Hz, based on the pn time resolution of 5.6718~ms, and only
exceeds the 90\% confidence detection limit in the region below 0.05~Hz.
\label{fig:timing1}}
\end{figure}

Part of this power spectrum is shown in Figure~\ref{fig:timing1}.  The power is
distributed as a $\chi^{2}$ probability distribution with 2 degrees of freedom (dof), 
and we used this fact along with the total number of trials, which is equal to the
number of frequency bins ($N_{trials}$ = 3,839,999 in this case) to determine a detection
threshold.  Here, the 90\% confidence detection limit is at a Leahy Power of 34.9, and
this is shown in Figure~\ref{fig:timing1}.  This limit is not exceeded at frequencies
above 0.005~Hz, but there are a large number of frequency bins below 0.005~Hz that 
exceed the limit.  We suspect that this may be related to low-frequency continuum 
noise; thus, we treat the $>$0.005~Hz case first.  

The highest power that we measure in the 0.005--88~Hz range is 33.4, which is just below 
the 90\% confidence detection threshold.  We do not consider this as even a marginal 
detection, but we use this value ($P_{max}$) to calculate an upper limit on the strength 
of a periodic signal.  As described in \cite{vdk89}, the 90\% confidence upper
limit is given by $P_{UL} = P_{max} - P_{exceed}$, where $P_{exceed}$ is the power level 
that is exceeded in 90\% of the frequency bins.  In our case, $P_{exceed} = 0.2$ so that
$P_{UL} = 33.2$, which, after converting to fractional rms units using the average
source and background count rates, corresponds to an upper limit on the rms noise 
level for a periodic signal of $<$2.3\%.  We also produced a power spectrum with 
$4.6\times 10^{-5}$~Hz frequency bins, but we still did not find any bins with power 
in the 0.005--88~Hz range with powers exceeding the 90\% confidence upper limit.

\begin{figure}
\plotone{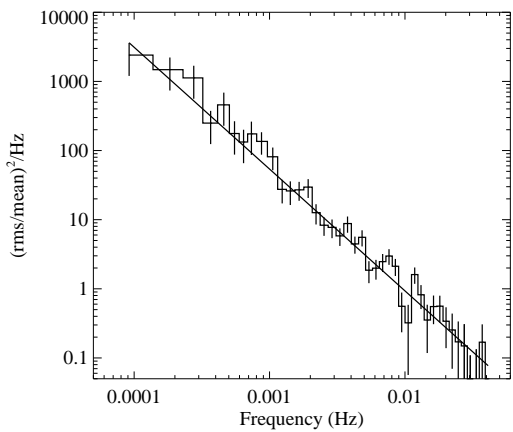}
\caption{Red noise rms-normalized power spectrum for IGR~J16207--5129.
The power-spectrum is fitted with a power-law (solid line).
\label{fig:timing2}}
\end{figure}

\begin{figure}
\plotone{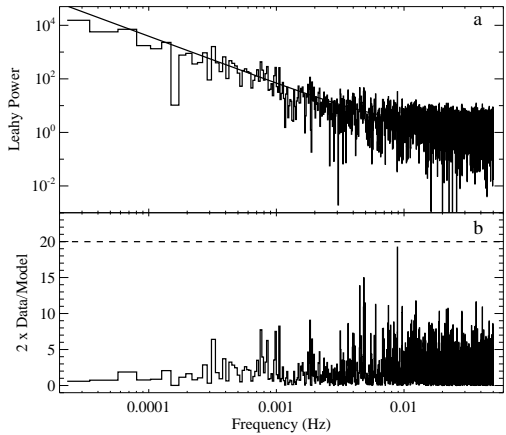}
\caption{(a) Leahy-normalized power spectrum for IGR~J16207--5129 with
a frequency binsize of $2.3\times 10^{-5}$~Hz.  The solid line is the
power-law model from the fit shown in Figure~\ref{fig:timing2}.  (b)
Two times the data-to-model ratio using the data and model shown in
panel a.  The horizontal dashed line shows the 90\% confidence detection
limit (after accounting for the number of trials).  The strongest 
signal is at 0.0089~Hz (112.56~s), but it is not statistically significant.
\label{fig:timing3}}
\end{figure}

To characterize the low-frequency noise, we produced a 0.4--15~keV pn light curve with
10~s time resolution, and made a new power spectrum with a Nyquist frequency of 0.05~Hz.
The new power spectrum consists of the average of power spectra from four $\sim$11~ks 
segments of the light curve, giving a minimum frequency of $9.2\times 10^{-5}$~Hz.  
In addition, we converted the power spectrum to rms normalization and rebinned the
power spectrum so that the power in each bin follows a Gaussian distribution, allowing
for us to fit a model to the power spectrum with $\chi^{2}$-minimization.  The resulting
power spectrum is shown in Figure~\ref{fig:timing2}, and the figure also shows that 
the power spectrum is well-described by a power-law model ($P = A (\nu/{\rm 1~Hz})^{-\alpha}$).
We obtain a power-law index of $\alpha = 1.76\pm 0.05$, and the integrated rms noise level
is 64\%$\pm$21\% (0.000092--0.05~Hz).  For the fit, the $\chi^{2} = 43.6$ for 39 dof, 
indicating that the power-law provides an acceptable description of the power spectrum.

To search for a periodic signal in the presence of red (i.e., power-law) noise requires
an extra step in the analysis.  Specifically, after producing another Leahy-normalized 
power spectrum, this time covering the 0.000023--0.05~Hz frequency range, multiplying
this power spectrum by 2 and dividing by the (re-normalized) power-law model leads to
a power spectrum where the power in each frequency bin follows a $\chi^{2}$ distribution
with 2 dof \citep{vdk89}.  This is illustrated by showing these power spectra
in Figure~\ref{fig:timing3}, and we can now search the powers in the bottom panel for
periodic signals.  After accounting for the number of trials (i.e., 2,176 frequency bins), 
the 90\% confidence detection limit is shown.  Although there are no signals that 
reach this detection limit, the maximum signal at 0.0089~Hz (112.56~s) is only slightly
below.  We do not consider this to be even a marginal detection, but we use the power
in this bin, $P_{max} = 19.2$ to determine the upper limit.  Using the same procedure 
as described above results in an upper limit on the rms noise level for a periodic 
signal of $<$1.7\%.

One caveat to the non-detection of a periodic signal is that it is possible for
the power from a periodic signal to be spread out in frequency by orbital motion.
This can happen if the duration of the observation is a substantial fraction of the
orbital period.  For IGR~J16207--5129, we do not know the orbital period; however, 
we do know that it is an HMXB and is expected to have an orbital period of between
several days and several weeks.  Thus, our $\sim$12~hr {\em XMM-Newton} observation 
should cover only a small fraction of the orbit.

\subsection{Energy Spectrum}

\subsubsection{Time-Averaged Spectrum}

We extracted pn, MOS1, and MOS2 energy spectra using the source and background regions
and filtering criteria described above, and we produced response matrices using the
SAS tools {\ttfamily rmfgen} and {\ttfamily arfgen}.  We obtained a total exposure time 
of 27~ks for the pn and 37~ks for each MOS unit.  The exposure time is lower for the 
pn due to a higher level of deadtime.   We rebinned the energy spectra by requiring at 
least 50 counts for each energy bin, leaving a pn spectrum with 722 bins, a MOS1 spectrum 
with 290 bins, and a MOS2 spectrum with 298 bins.  Although we use the 0.4--15~keV
energy band for pn light curves, we restrict the spectral analysis to 0.4--12~keV for
the pn and MOS units as the calibration\footnote{A document on the pn and EPIC calibration 
dated 2008 April by M.~Guainazzi can be found at 
http://xmm2.esac.esa.int/docs/documents/CAL-TN-0018.pdf.} has focused on this energy range.  

\begin{figure}
\plotone{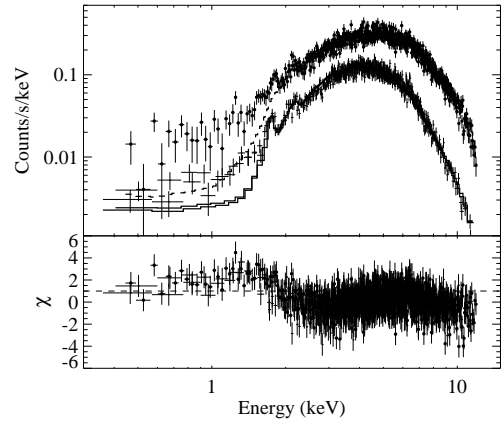}
\vspace{-0.8cm}
\caption{{\em XMM-Newton} pn and MOS energy spectrum for IGR~J16207--5129 fitted 
with an absorbed power-law.  Residuals (lower panel) illustrate the presence of
a soft excess.  The pn points are marked with points while the
MOS1 and MOS2 points are not (upper and lower panels), and the line used for the 
pn model is dashed while the MOS model lines are solid (upper panel).
\label{fig:spectrum_counts}}
\end{figure}

\begin{figure}
\plotone{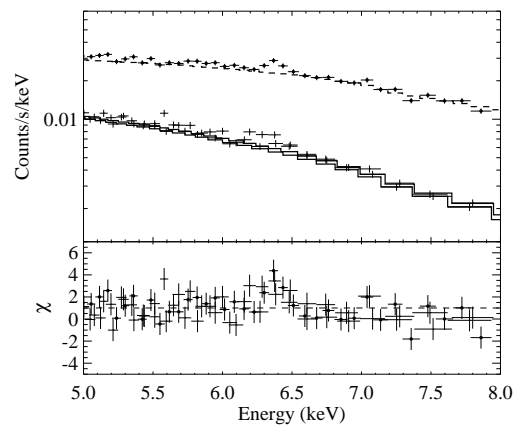}
\vspace{-0.8cm}
\caption{{\em XMM-Newton} pn and MOS energy spectrum in the iron line region
for IGR~J16207--5129 fitted with an absorbed power-law.  Residuals (lower panel) 
illustrate the presence of an iron line.  The pn points are marked with points 
while the MOS1 and MOS2 points are not (upper and lower panels), and the line 
used for the pn model is dashed while the MOS model lines are solid (upper panel).
\label{fig:counts_iron}}
\end{figure}

We used the XSPEC software package to jointly fit the 3 spectra with an absorbed 
power-law model.  To account for absorption, we used the photoelectric absorption 
cross sections from \cite{bm92} and elemental abundances from \cite{wam00}, which 
correspond to the estimated abundances for the interstellar medium.  We left the
relative normalization between the instruments as free parameters, and a
$\chi^{2}$-minimization fit indicates that the MOS normalizations are slightly 
higher (3\%$\pm$1\%) than the pn normalization.  The absorbed power-law model,
with the best fit values of $N_{\rm H} = 9\times 10^{22}$~cm$^{-2}$ for the column
density and $\Gamma = 0.97$ for the power-law photon index, appears to provide a 
good description of the spectral continuum above 2~keV; however, the fit is not
statistically acceptable overall with $\chi^{2}/\nu = 1778/1305$.  The counts 
spectrum and residuals for the absorbed power-law fit are shown in 
Figure~\ref{fig:spectrum_counts}.  Large residuals are seen below 2~keV, suggesting 
the presence of a soft excess.  We also produced pn and MOS spectra with a higher
level of binning to study the iron K$\alpha$ region of the spectrum.  The 
spectrum and residuals shown in Figure~\ref{fig:counts_iron} indicates the presence
of an iron emission line near 6.4~keV.

We refit the spectrum (going back to the lower level of binning) after adding a 
Gaussian to model the iron line and a second spectral component to account for 
the soft excess.  We tried different models for the second spectral component, 
including a Bremsstrahlung model, a black-body, and a power-law.  In each case, 
the second component was absorbed, but we fixed $N_{\rm H}$ for this component 
to the Galactic value, $1.7\times 10^{22}$ cm$^{-2}$ \citep{dl90}.  We also used 
this level of absorption for the iron line.  Adding the iron line and using a
Bremsstrahlung model for the second component yields a much improved fit (over 
the power-law alone) of $\chi^{2}/\nu = 1402/1300$.  The best fit Bremsstrahlung
temperature is 189~keV, which is approaching the upper end of the range allowed 
by the XSPEC model {\tt bremss}.  We derive a 90\% confidence lower limit on the 
temperature of 50~keV, indicating that if the soft excess is due to Bremsstrahlung
emission, it is rather hot.  While it is tempting to take the high Bremsstrahlung
temperature as an indication that the soft excess is non-thermal, using a 
black-body model for the second component yields a fit of comparable quality,
$\chi^{2}/\nu = 1403/1300$, and the black-body temperature is $\sim$0.6~keV.  
Finally, if a power-law is used for the second component, the quality of the
fit is identical to the Bremsstrahlung fit, $\chi^{2}/\nu = 1402/1300$, and 
the power-law photon index is $\Gamma = 0.9^{+0.5}_{-0.4}$.  Although the 
statistical quality of the fits does not allow us to determine which model is 
the best to use for the second, soft excess, component, the fact that the 
power-law has a photon index that is consistent with the value of $\Gamma$ 
found for the primary power-law component, $\Gamma = 1.15^{+0.07}_{-0.05}$
(90\% confidence errors), suggests that it may be possible to interpret 
the soft excess as an unabsorbed portion of the primary component, and this
possibility is explored further in \S$3.2.2$.  Thus, we proceed by focusing 
on the model that uses the power-law for the second component.

\begin{figure}
\plotone{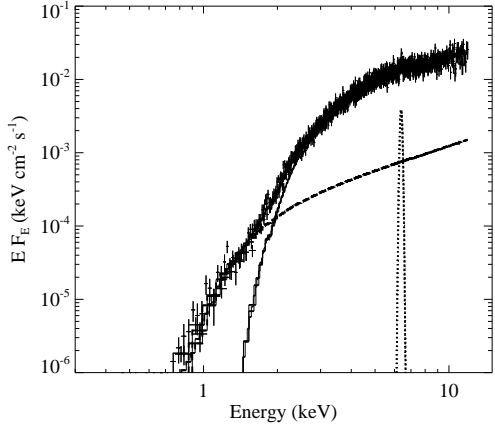}
\caption{{\em XMM-Newton} pn and MOS unfolded energy spectrum for IGR~J16207--5129
with a model consisting of an absorbed power-law (dashed line), a less absorbed power-law 
to account for the soft excess (dotted line), and an iron K$\alpha$ emission line (solid
line).
\label{fig:spectrum_efe}}
\end{figure}

The different components of the model with an absorbed power-law, a power-law 
with just interstellar absorption, and an iron line are shown in 
Figure~\ref{fig:spectrum_efe}, and the model parameters are shown in 
Table~\ref{tab:spectra}.  The primary power-law component has the value of
$\Gamma$ given above absorbed by a column density of $N_{\rm H} = 
(1.19^{+0.06}_{-0.05})\times 10^{23}$ cm$^{-2}$, and the 0.5--10~keV unabsorbed
flux of the power-law component is $3.7\times 10^{-11}$ ergs~cm$^{-2}$~s$^{-1}$, 
which is a factor of $\sim$20 higher than the soft excess component.  The 
parameters of the narrow iron K$\alpha$ line are well-constrained.  The energy 
of the line is $6.39\pm 0.03$~keV, which is consistent with iron ionization states
between neutral (FeI) and $\sim$FeX \citep{nagase86}.  The 90\% confidence upper 
limit on the width of the line is $<$0.12~keV.  The line is clearly detected, 
but it is not extremely strong with an equivalent width of $42\pm 12$~eV after 
correcting the line and power-law components for absorption.

\subsubsection{Spectral Properties vs.~Intensity}

\begin{figure}
\plotone{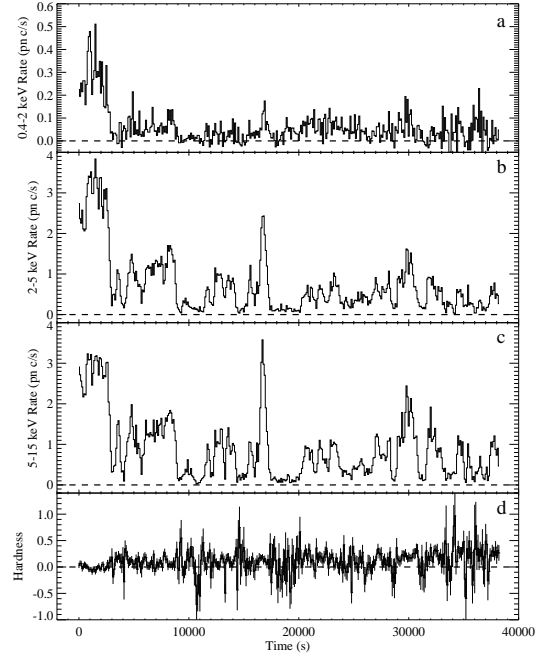}
\caption{IGR~J16207--5129 light curves in 0.4--2~keV (a), 2--5~keV (b), and 
5--15~keV (c) energy bands and the hardness vs. time (d).  Defining the
5--15~keV rate as $C_{hard}$ and the 2--5~keV rate at $C_{soft}$, the hardness 
is ($C_{hard}$--$C_{soft}$)/($C_{hard}$+$C_{soft}$).  The time resolution in each
case is 100~s, and we have subtracted the background contribution.\label{fig:ht}}
\end{figure}

A caveat on the above spectral analysis is that the parameters represent average 
spectral properties for a highly variable source.  As an initial look at how the 
spectrum changes with flux level, we extracted deadtime corrected and background 
subtracted pn light curves with 100~s time resolution in the 0.4--2~keV, 2--5~keV,
and 5--15~keV energy bands.  Figure~\ref{fig:ht} shows these light curves (panels
a--c) along with the hardness vs.~time (panel d).  The hardness is calculated
using the rates in the 2--5~keV and 5--15~keV energy bands (defined as $C_{soft}$ 
and $C_{hard}$, respectively) according to ($C_{hard}$--$C_{soft}$)/($C_{hard}$+$C_{soft}$).
The most striking aspect of Figure~\ref{fig:ht} is how similar the 2--5~keV and
5--15~keV light curves appear, suggesting that the variability is relatively
independent of energy.  Although the statistics are poorer for the 0.4--2~keV
light curve, many of the features in this light curve can be associated with
variability in the higher energy light curves.  A close inspection of the hardness
(Figure~\ref{fig:ht}d) shows that the light curves do have some energy dependence.
While there are some exceptions, for most of the deepest dips, the spectra are 
softer, with the hardness being less than zero for many dips.  In contrast, most
flares are harder, with typical hardness values of $\sim$0.2--0.3.  While the 
sharp flare that occurs at a time of $\sim$16,000~s follows the typical behavior, 
a notable exception is found for the flare seen in the first $\sim$3,000~s of the
observation, which has hardness values between --1 and $\sim$0.

\begin{figure}
\plotone{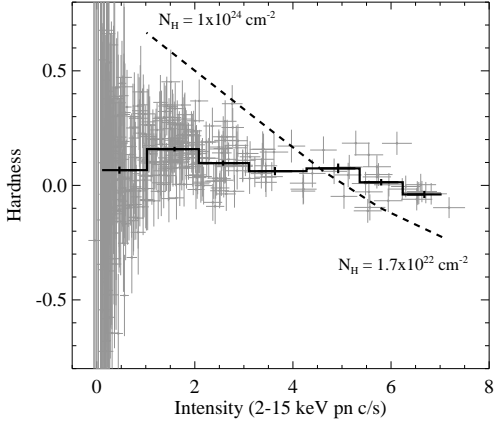}
\caption{Hardness-Intensity diagram for IGR~J16207--5129.  We used the 100~s time 
resolution light curves for the counts in the 5--15~keV band ($C_{hard}$) and the 
2--5~keV band ($C_{soft}$).  The hardness is defined as 
($C_{hard}$--$C_{soft}$)/($C_{hard}$+$C_{soft}$).  In addition to the individual
100~s points, which are grey in color, the averages in seven intensity bins are 
shown (the black histogram).  The dashed line shows the hardness-intensity 
relationship expected if the change in count rate was due only to a change in 
column density.\label{fig:hi}}
\end{figure}

The same hardness values are plotted vs.~the 2--15~keV count rate (i.e., intensity)
in Figure~\ref{fig:hi}, with each point corresponding to a 100~s time bin.  In 
addition, we calculated weighted averages in seven intensity bins, and these are also 
shown in the figure.  At the high intensity end, the average hardness is dominated 
by the softer flare at the beginning of the observation.  Then, the spectrum is 
harder at intermediate intensities before softening again below 1 c/s.  We 
calculated the expected hardness change if the variability is due only to a 
change in $N_{\rm H}$, and the prediction as $N_{\rm H}$ changes from 
$1.7\times 10^{22}$ cm$^{-2}$ to $10^{24}$ cm$^{-2}$ is shown in Figure~\ref{fig:hi}.  
The predicted change in hardness is drastically larger than the change we observe, 
indicating that changing $N_{\rm H}$ cannot be the sole cause of the variability.  

\begin{figure}
\plotone{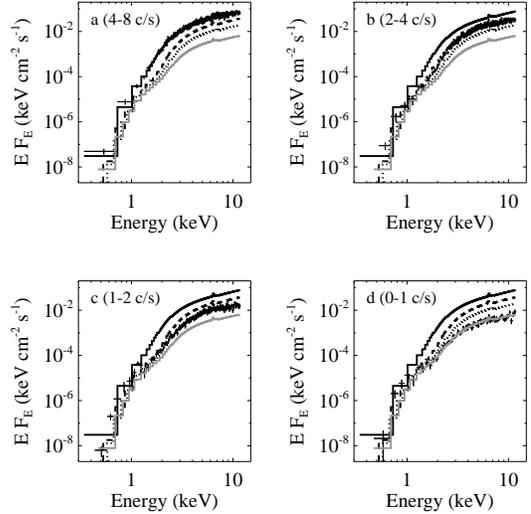}
\caption{Unfolded pn energy spectra after dividing the data into the following
different intensity levels:  (a) 4--8 pn c/s; (b) 2--4 pn c/s; (c) 1--2 pn c/s;
and (d) 0--1 pn c/s.  In each panel, the best fit partial covering fraction 
models are plotted for all four intensity levels.  
\label{fig:efe_intensity}}
\end{figure}

To investigate which spectral parameters do change with intensity, we used the 
0.4--15~keV pn light curve to divide the pn and MOS data into spectra at four 
intensity levels:  4--8 pn c/s, 2--4 pn c/s, 1--2 pn c/s, and 0--1 pn c/s.  We 
used all of the data from the first 39~ks of the observation and obtained pn 
exposure times of 2,485~s, 6,177~s, 8,520~s, and 9,940~s and MOS exposure times 
of 3,412~s, 8,482~s, 11,700~s, and 13,650~s for the four intensity levels, 
respectively.  In fitting these spectra, we simplified the two power-law model 
by requiring the same $\Gamma$ for both power-laws, which is equivalent to using 
a model where a fraction, $f$, of the power-law component is absorbed both by 
interstellar material and by the extra material local to the source, while a 
fraction, 1--$f$ is absorbed only by interstellar material.  Table~\ref{tab:intensity}
shows the parameters obtained for the total spectrum and also for each of the
four intensity levels, and Figure~\ref{fig:efe_intensity} shows the unfolded
spectra for the four levels.  While we performed all of the fits with pn and 
MOS spectra, Figure~\ref{fig:efe_intensity} shows only the pn spectra for
clarity.

One interesting result that we obtain from the spectral parameters shown in
Table~\ref{tab:intensity} is that, at $N_{\rm H} = (6.4\pm 0.4)\times 10^{22}$
cm$^{-2}$, the column density for the highest intensity level is significantly
less that for the lower levels.  This spectrum is dominated by the flare that
occurs during the first 3,000~s of the observation, and this indicates that 
lower $N_{\rm H}$ is the reason that this flare is softer.  The parameters also
indicate that there are two reasons why the spectrum is softer during the 
deepest dips.  First, the power-law index indicates that the source gradually
softens toward the lower levels, going from $\Gamma = 1.06\pm 0.05$ at 4--8 c/s
to $1.30\pm 0.12$ at 0--1 c/s.  Secondly, the parameters and the spectra shown
in Figure~\ref{fig:efe_intensity} indicate that the flux of the soft excess
does not decrease as much as the flux of the primary power-law component.  
Between 2--4 c/s and 0--1 c/s, the flux of the primary power-law component
changes by a factor of $5.1\pm 0.5$ while the flux of the soft excess changes
by a factor of $2.0\pm 0.5$.  The iron line parameters show that the equivalent
width of the iron line is consistent with being constant while the flux of the
line changes with the continuum flux.

\section{Discussion}

\subsection{Is IGR J16207--5129 an Obscured HMXB?}

While the {\em Chandra} position and the optical/IR spectroscopy leave essentially
no doubt that IGR~J16207--5129 is an HMXB, it has not been entirely clear whether
the column density is high enough to require that part of the absorbing material
is local to the source or not.  The only other soft X-ray measurement besides the
one reported here is the {\em Chandra} observation, and \cite{tomsick06} reported
a value of $N_{\rm H} = (3.7^{+1.4}_{-1.2})\times 10^{22}$ cm$^{-2}$ (90\% confidence
errors), while many of the obscured HMXBs have column densities in excess of 
$10^{23}$ cm$^{-2}$.  When one considers atomic and molecular hydrogen, the total 
Galactic column density is near $2.4\times 10^{22}$ cm$^{-2}$ \citep{dl90,dht01,tomsick08a}, 
so the {\em Chandra} measurement is only marginally higher than the Galactic value.

However, the value that we measure with {\em XMM-Newton} is substantially higher, 
$N_{\rm H} = (1.19^{+0.06}_{-0.05})\times 10^{23}$ cm$^{-2}$, which is significantly in 
excess of the Galactic value.  To check on whether the difference between the 
{\em Chandra} and {\em XMM-Newton} measurements is related to variability in the column 
density, we re-analyzed the {\em Chandra} spectrum.  In our previous {\em Chandra} 
analysis \citep{tomsick06}, we used the \cite{ag89} rather than \cite{wam00} abundances, 
and we fitted the spectrum with only a single power-law.  First, we find that when 
we use the \cite{wam00} abundances and re-fit the {\em Chandra} spectrum, we 
obtain a value of $N_{\rm H} = (5.4^{+2.1}_{-1.7})\times 10^{22}$ cm$^{-2}$, which is
already somewhat higher than found in the previous analysis.  Second, the quality
of the {\em Chandra} spectrum does not allow for the detection of the soft excess, 
but we re-fitted the {\em Chandra} spectrum after including a soft excess with
the values of $N_{\rm H}$ and $\Gamma$ measured with {\em XMM-Newton} (see 
Table~\ref{tab:spectra}).  This fit gives values for the primary power-law component
of $\Gamma = 0.7\pm 0.7$ and $N_{\rm H} = 7.9^{+4.4}_{-3.2}\times 10^{22}$ cm$^{-2}$, 
which are both consistent with the values measured with {\em XMM-Newton}.  

Thus, the {\em Chandra} spectrum allows for the possibility that IGR~J16207--5129 is 
an obscured HMXB while the {\em XMM-Newton} spectrum constrains $N_{\rm H}$ to be 
significantly higher than the Galactic value, requiring that the source is an obscured 
HMXB.  With {\em XMM-Newton}, we see that the $N_{\rm H}$ can change, with the first
3,000~s of the observation having $N_{\rm H} = (6.4\pm 0.4)\times 10^{22}$ cm$^{-2}$,
while the $N_{\rm H}$ was about a factor of two higher for the rest of the observation.
The {\em Chandra} spectrum is consistent with either level.  Even at the highest
$N_{\rm H}$ seen by {\em XMM-Newton}, the amount of local absorbing material is 
certainly not as much as seen in some of the most extreme systems like 
IGR~J16318--4848, which has $N_{\rm H} = 2\times 10^{24}$ cm$^{-2}$.  This is consistent 
with optical and IR observations which have shown P Cygni profiles, forbidden emission 
lines (indicating a supergiant B[e] spectral type), and IR excesses from local material 
for IGR~J16318--4848 \citep{fc04,moon07,rahoui08} but not for IGR~J16207--5129 
\citep{ns07,nespoli08,rahoui08}.

\subsection{Comparison to other HMXBs Lacking Pulsations}

With the results of our timing study, IGR~J16207--5129 joins a relatively short
list of accreting HMXBs that show very hard energy spectra, resembling known 
neutron star HMXBs, but that do not seem to exhibit pulsations (at least in the 
expected frequency range).  One example is 4U~1700--377, which has an extremely 
massive and luminous O6.5~Iaf+ companion in a 3.41~day orbit with the compact 
object.  The properties of this source as seen by {\em XMM-Newton} are remarkably
similar to those of IGR~J16207--5129 \citep{vandermeer05}.  Light curves with 
10~s time resolution show flares and dips where the flux changes by factors of 
at least 5 on ks time scales.  While the spectral analysis was limited to the
fainter time periods because of photon pile-up, spectral fits show a direct
component with a hard power-law index ($\Gamma = 1.08$--1.87 for the ``low-flux'' 
spectrum, depending on the exact model used) and a high level of absorption 
($N_{\rm H} = 6.8\times 10^{22}$~cm$^{-2}$ to $N_{\rm H} = 2.0\times 10^{23}$~cm$^{-2}$)
along with a soft excess.  While the light curves for 4U~1700--377 and
IGR~J16207--5129 are somewhat unusual in their extreme variability on long
time scales, it should be noted that many pulsating HMXBs (e.g., Vela~X-1
and GX~301--2) also show very similar energy spectra to those described here, 
with hard primary power-law components and soft excesses \citep{nagase89}.

A comparison of the power spectra of 4U~1700--377 and IGR~J16207--5129 show
similarities also.  The typical power spectrum of an HMXB pulsar has 3 parts:
A relatively flat portion at frequencies below the pulsation frequency; 
a region where the pulsations, harmonics, and sometimes quasi-periodic 
oscillations (QPOs) are observed, and a higher frequency region where 
the power spectrum can be described as a power-law with a slope of 1.4--2.0
\citep{bh90,ht93,chakrabarty01}.  However, a uniform study of 12 HMXBs 
with {\em EXOSAT} indicated that 4U~1700--377 (and, interestingly, Cyg X-3)
deviated from this pattern by showing only a steep power-law in the
0.002--1~Hz frequency range \citep{bh90}, like IGR~J16207--5129.  \cite{bh90} 
quote 0.008--29~Hz integrated rms noise levels for 4U~1700--377 between 6 and 
12\%, and if we recalculate the IGR~J16207--5129 noise level for this 
frequency range, we obtain 11.7\%$\pm$2.7\%.

While there are many similarities between these two sources, one possible 
difference is that the average luminosity of 4U~1700--377 is probably
higher.  To avoid model-dependent flux measurements, we compare the average
20--40~keV flux measurements made by {\em INTEGRAL}.  For 4U~1700--377, 
\cite{bird07} quote a value of $208.1\pm 0.1$ millicrab compared to
$3.3\pm 0.1$ millicrab for IGR~J16207--5129.  The distance to 4U~1700--377
is estimated at 1.9~kpc \citep{ankay01}, so the ratio of fluxes for the
two sources would imply a distance of $\sim$15~kpc for IGR~J16207--5129 if
their 20--40~keV luminosities were the same, which is at the very upper 
limit of the distance range derived by \cite{nespoli08}.  At the best
estimate for the IGR~J16207--5129 distance, 6.1~kpc, the luminosity of 
IGR~J16207--5129 would be about 6 times lower than 4U~1700--377.  At a 
distance of 6.1~kpc, the average IGR~J16207--5129 20--40 keV luminosity is 
$1.1\times 10^{35}$ ergs~s$^{-1}$, and the average 0.5--10 keV unabsorbed 
luminosity is $1.6\times 10^{35}$ ergs~s$^{-1}$.

Despite soft X-ray spectra that are similar to neutron star HMXBs, the 
nature of the compact object is still unclear for both 4U~1700--377 and
IGR~J16207--5129 because of the lack of X-ray pulsations.  One reason that
the black hole possibility has been taken seriously for 4U~1700--377 is
that the compact object mass has been measured to be 
$2.44\pm 0.27$~\Msun~\citep{clark02}, which is significantly higher than 
the values close to 1.4\Msun~measured for most neutron stars \citep{tc99},
and could indicate that 4U~1700--377 harbors a low mass black hole.  
Obtaining a compact object mass measurement for IGR~J16207--5129 would 
certainly be interesting and may be feasible as it is relatively bright 
in the optical and IR (see \S$1$).  However, the mass measurement would be 
challenging because accurate spectroscopy would be required to measure the 
massive B1~Ia companion's radial velocity curve.  Rather than black holes, 
these sources may harbor very slowly rotating neutron stars.  Several such 
X-ray binaries are known:  IGR~J16358--4726 with its 1.64~hr pulsations
\citep{patel04}; 2S~0114+650 with its 2.7~hr pulsations and B1~Ia spectral
type \citep[][and references therein]{farrell08}, which is the same spectral
type as IGR~J16207--5129; 4U~1954+319 with its likely 5~hr pulsations 
\citep{mattana06}; and 1E 161348--5055.1 with its likely 6.7~hr pulsations
\citep{deluca06}.  With our $\sim$12~hr {\em XMM-Newton} observation of 
IGR~J16207--5129, we can rule out pulsation periods in the $\sim$1--2 hr
range but probably not in the $\gsim$4 hr range.  

Another HMXB with a neutron star-like spectrum that has been well-studied 
without detecting pulsations is 4U~2206+54.  This system has an O9.5~V or 
O9.5~III companion \citep{ribo06} and a 9.6 or 19.25 day orbital period
\citep{corbet07}.  Like IGR~J16207--5129, its X-ray emission is highly
variable \citep{masetti04}, but its energy spectrum is significantly 
different, showing no evidence for local absorption.  Its $10^{-3}$--1 Hz
power spectrum is dominated by strong red (i.e., power-law) noise 
\citep{nr01,torrejon04,blay05}, which is similar to IGR~J16207--5129 and 
4U~1700--377.  However, for 4U~2206+54, the possible detection of a
cyclotron line at $\sim$30~keV provides some additional evidence that
the compact object is a neutron star \citep{torrejon04,blay05}.  If
confirmed, this would indicate that the lack of pulsations and the red
noise power spectrum should not be taken as evidence for a black hole.
Rather, the non-detections of pulsations could be related to a system
geometry where, e.g., the neutron star magnetic and spin axes are nearly 
aligned, or the non-detections could be due to spin periods that are
longer than the duration of the observations.

Finally, it is worth noting that several IGR HMXBs besides IGR~J16207--5129 
have not yet shown X-ray pulsations.  IGR~J19140+0951 has similarities to
IGR~J16207--5129, with a B0.5~I spectral type, a high level of obscuration
(although the level depends on orbital phase), and a soft excess 
\citep[see][and references therein]{prat08}.  IGR~J16318--4848 (see
\S$4.1$) also has not exhibited pulsations despite a high level of X-ray
variability that has been seen in several soft X-ray observations.
Although the apparent lack of pulsations may be due to the fact that the 
timing properties of these sources have not been very well-studied yet, 
it is possible that the IGR HMXBs include a population of systems with 
black holes or very slowly rotating neutron stars.

\subsection{Soft Excess and Iron Line}

The presence of a soft excess and emission lines are very common in both pulsating
and non-pulsating obscured HMXBs.  In addition to iron lines, some sources (e.g., 
Vela~X-1 and 4U~1700--377) have emission lines from Si, Ne, etc.~in the 0.5--3~keV
range, making up at least part of the soft excess \citep{watanabe06,boroson03,vandermeer05}.  
Although our spectral analysis does not allow us to definitively rule out other 
models besides a power-law (e.g., Bremsstrahlung or black-body) for the soft excess,
the fact using a power-law leads to a power-law index that is consistent with the
value of $\Gamma$ for the primary power-law components allows for two possible 
physical interpretations.  One possibility is the partial covering of the power-law
source by absorbing material (likely the stellar wind), which is the scenario envisioned 
for the spectral fits described in \S$3.2.2$.  Another possible explanation is that 
the soft excess emission originates as part of the primary power-law component, but 
it is scattered in the stellar wind.  In this picture, the fact that the soft 
excess has a much lower column density than the primary component would indicate
that the photons that are part of the soft excess come from the edge (i.e., within 
one optical depth of the edge) of the stellar wind.  

Although it is not clear whether any of the IGR~J16207--5129 soft excess comes
from emission lines, the fact that the flux of the Fe~K$\alpha$ line at 6.4~keV 
correlates with the overall source flux (i.e., the equivalent width is consistent
with being constant) is different from Vela~X-1 and 4U~1700--377, which both show
lines with very large equivalent widths at low luminosities.  Regardless of the 
presence (or not) of lower energy lines in IGR~J16207--5129, the energy of the 
iron line indicates a low ionization state implies that the line originates in 
cool material.  One possible location for the cool material is an accretion disk 
around the IGR~J16207--5129 compact object; however, many HMXBs with strong stellar 
winds have compact objects that accrete directly from the wind with little or no 
disk accretion.  The equivalent width of $42\pm 12$~eV that we measure for the 
IGR~J16207--5129 iron line is consistent with simulations in which a spherical 
distribution of cold matter with solar abundances and 
$N_{\rm H}\sim 10^{23}$~cm$^{-2}$ is illuminated by a central X-ray source 
\citep{matt02}, and this is also true for several other IGR HMXBs \citep{walter06}.
Although it is not obvious that the matter in a stellar wind will be cold
(i.e., neutral), it has been suggested that the wind might be a clumpy two-phase 
medium with cool, dense regions that are responsible for the emission lines along 
with hot, highly-ionized regions \citep[][and references therein]{vandermeer05}.  
The partial covering model that we use to fit the IGR~J16207--5129 continuum 
would also be consistent with absorption of the X-ray source by a clumpy wind.

\section{Summary and Conclusions}

The {\em XMM-Newton} observation of IGR~J16207--5129 confirms that it is a 
member of the group of obscured IGR HMXBs.  We measure a column density of
$N_{\rm H} = (1.19^{+0.06}_{-0.05})\times 10^{23}$ cm$^{-2}$ for the average
spectrum.  We find that the column density could have been that high during 
the only previous soft X-ray observation of this source with {\em Chandra}, 
but a detailed spectral analysis of the {\em XMM-Newton} data shows that 
$N_{\rm H}$ can vary by a factor of $\sim$2.  We detect an iron line in
the energy spectrum, and its strength is consistent with what is expected
for a spherical distribution of material with the measured $N_{\rm H}$ around 
the compact object.  Although the X-ray spectrum is similar to those seen
from other neutron star HMXBs with a hard primary power-law component and
a soft excess, we do not detect the pulsations that might be expected if
the compact object is a neutron star.  

A detailed comparison between IGR~J16207--5129 and another apparently
non-pulsating HMXB, 4U~1700--377, shows strong similarities in spectral 
and timing properties.  Most notably, the power spectra of both sources 
can be described as a single power-law down to $10^{-3}$-$10^{-4}$~Hz.  
Since pulsating HMXBs show power spectra that break near the pulsation
frequency, it is possible that both of these sources harbor very slowly
rotating neutron stars (although the possibility that the compact object
is a black hole cannot be ruled out entirely).  We note that several 
of the IGR HMXBs are either known to harbor slowly rotating neutron stars
or may harbor slowly rotating neutron stars in cases where pulsations have 
not been detected.  It is possible that in addition to uncovering obscured
HMXBs, the HMXBs that {\em INTEGRAL} is uncovering tend to contain slow
rotators.

\acknowledgments

JAT acknowledges partial support from National Aeronautics and Space Administration 
(NASA) {\em XMM-Newton} Guest Observer award number NNX07AQ11G.  JAT thanks Nora 
Loiseau of the {\em XMM-Newton} User Support Group, Joern Wilms, and Peter Woods 
for helpful information concerning the {\em XMM-Newton} data analysis.  We thank
an anonymous referee for useful suggestions on the spectral analysis.



\clearpage

\begin{table}
\caption{Spectral Results\label{tab:spectra}}
\begin{minipage}{\linewidth}
\footnotesize
\begin{tabular}{cc} \hline \hline
Parameter & Value\footnote{The errors on the parameters are for $\Delta\chi^{2} = 2.7$, 
corresponding to 90\% confidence for one parameters of interest.}\\ \hline
\multicolumn{2}{c}{Primary Power-Law Component}\\ \hline
$N_{\rm H}$ & $(1.19^{+0.06}_{-0.05})\times 10^{23}$ cm$^{-2}$\\
$\Gamma$ & $1.15^{+0.07}_{-0.05}$\\
$F_{PL}$ (unabsorbed, 0.5--10 keV) & $(3.68\pm 0.10)\times 10^{-11}$ ergs cm$^{-2}$ s$^{-1}$\\ \hline
\multicolumn{2}{c}{Soft Excess}\\ \hline
$N_{\rm H}$ & $1.7\times 10^{22}$ cm$^{-2}$ (fixed)\\
$\Gamma$ & $0.9^{+0.5}_{-0.4}$\\
$F_{PL}$ & $(1.8^{+1.0}_{-0.7})\times 10^{-12}$\\ \hline
\multicolumn{2}{c}{Iron Line Parameters}\\ \hline
$N_{\rm H}$ & $1.7\times 10^{22}$ cm$^{-2}$ (fixed)\\
$E_{\rm line}$ & $6.39\pm 0.03$ keV\\
$\sigma_{\rm line}$ & $<$0.12 keV\\
$N_{\rm line}$ & $(1.6^{+0.5}_{-0.4})\times 10^{-5}$ photons cm$^{-2}$ s$^{-1}$\\
Equivalent Width\footnote{This is the equivalent width for the unabsorbed Gaussian line relative to the unabsorbed power-law continuum components.} &  $42\pm 12$ eV\\ \hline
$\chi^{2}/\nu$ & 1402/1300\\ \hline
\end{tabular}
\end{minipage}
\end{table}

\begin{table}
\caption{Spectral Results with the Partial Covering Model vs.~Intensity\label{tab:intensity}}
\begin{minipage}{\linewidth}
\footnotesize
\begin{tabular}{cccccc} \hline \hline
          & Value       & Value      & Value      & Value      & Value\\
Parameter & (all rates) & (4--8 c/s) & (2--4 c/s) & (1--2 c/s) & (0--1 c/s)\\ \hline
\multicolumn{6}{c}{Partially Absorbed Power-law Parameters}\\ \hline
$N_{\rm H}$ ($\times 10^{23}$ cm$^{-2}$) &  $1.00^{+0.04}_{-0.05}$ & $0.64\pm 0.04$ & $1.14\pm 0.06$ & $1.31\pm 0.09$ & $1.24\pm 0.13$\\
$\Gamma$ & $1.13\pm 0.02$ & $1.06\pm 0.05$ & $1.15\pm 0.06$ & $1.21\pm 0.08$ & $1.30\pm 0.12$\\
$f$\footnote{Fraction of the power-law flux absorbed by interstellar material and material local to the source.  A fraction equal to 1--$f$ is absorbed by just interstellar material (at $N_{\rm H} = 1.7\times 10^{22}$ cm$^{-2}$), leading to the soft excess seen in the spectrum.} & $0.963\pm 0.003$ & $0.964\pm 0.008$ & $0.980\pm 0.004$ & $0.971\pm 0.005$ & $0.951^{+0.009}_{-0.011}$\\
$F_{PL}$\footnote{Unabsorbed 0.5--10~keV flux in units of $10^{-11}$ ergs~cm$^{-2}$~s$^{-1}$.} & $3.84^{+0.07}_{-0.04}$ & $11.2\pm 0.3$ & $6.06^{+0.24}_{-0.11}$ & $3.33^{+0.18}_{-0.17}$ & $1.22^{+0.12}_{-0.10}$\\
$F_{PPL} = F_{PL}f$\footnote{$F_{PPL}$ is the unabsorbed 0.5--10~keV flux of the primary power-law component in units of $10^{-11}$ ergs~cm$^{-2}$~s$^{-1}$ and is derived from $F_{PL}$ and $f$.} & $3.70^{+0.07}_{-0.04}$ & $10.8\pm 0.3$ & $5.94^{+0.24}_{-0.11}$ & $3.23\pm 0.17$ & $1.16^{+0.11}_{-0.10}$\\
$F_{SE} = F_{PL}(1-f)$\footnote{$F_{SE}$ is the unabsorbed 0.5--10~keV flux of the soft excess component in units of $10^{-11}$ ergs~cm$^{-2}$~s$^{-1}$ and is derived from $F_{PL}$ and $f$.} & $0.142\pm 0.012$ & $0.40\pm 0.09$ & $0.12\pm 0.02$ & $0.10\pm 0.02$ & $0.060\pm 0.013$\\ \hline
\multicolumn{6}{c}{Iron Line Parameters}\\ \hline
$E_{\rm line}$ (keV) & $6.39\pm 0.03$ & $6.44\pm 0.08$ & $6.33\pm 0.05$ & $6.41^{+0.07}_{-0.06}$ & $6.39^{+0.10}_{-0.09}$\\
$\sigma_{\rm line}$ (keV) & $0.06^{+0.03}_{-0.06}$ & 0.06\footnote{Fixed.} & 0.06$^{e}$ & 0.06$^{e}$ & 0.06$^{e}$\\
$N_{\rm line}$\footnote{Iron line flux in units of $10^{-5}$ photons~cm$^{-2}$~s$^{-1}$.} & $1.6^{+0.5}_{-0.7}$ & $4.9^{+2.1}_{-3.3}$ & $2.5^{+0.8}_{-1.1}$ & $1.0\pm 0.6$ & $0.48^{+0.36}_{-0.32}$\\
EW (eV) & $42^{+13}_{-18}$ & $44^{+18}_{-30}$ & $43^{+14}_{-19}$ & $32\pm 19$ & $44^{+33}_{-29}$\\ \hline
$\chi^{2}/\nu$ & 1402/1301 & 489/468 & 528/503 & 432/397 & 232/193\\ \hline
\end{tabular}
\end{minipage}
\end{table}

\end{document}